\documentclass[apl,twocolumn,showpacs,floatfix,10pt]{revtex4}%
\usepackage{amsfonts}
\usepackage{amsmath}
\usepackage{amssymb}
\usepackage{graphicx}
\usepackage{epsfig}%
\begin{document}

\title{Quantised Vortices and Four-Component Superfluidity of Semiconductor Excitons}

\author{Romain Anankine$^{1,\star}$, Mussie Beian$^{1,2,\star}$, Suzanne Dang$^1$, Mathieu Alloing$^{1}$, Edmond Cambril$^3$, Kamel Merghem$^{3}$, Carmen Gomez Carbonell$^3$, Aristide Lema\^{i}tre$^3$ and Fran\c{c}ois Dubin$^{1,2}$}

\affiliation{$^1$ UPMC Univ Paris 06, CNRS-UMR 7588, Institut des NanoSciences de Paris, 4 Place Jussieu, F-75005 Paris, France}

\affiliation{$^2$ ICFO-The Institute of Photonic Sciences, 
Av. Carl Friedrich Gauss, num. 3, 08860 Castelldefels, Spain}

\affiliation{$^3$ Laboratoire de Photonique et Nanostructures, LPN/CNRS, Route de
Nozay, 91460 Marcoussis, France}

\affiliation{$^*$ contributed equally}

\begin{abstract}
We study spatially indirect excitons of GaAs quantum wells, confined in a 10 $\mu$m electrostatic trap. Below a critical temperature of about 1 Kelvin, we detect macroscopic spatial coherence and quantised vortices in the weak photoluminescence emitted from the trap. These quantum signatures are restricted to a narrow range of density, in a dilute regime. They manifest the formation of a four-component superfluid, made by a low population of optically bright excitons coherently coupled to a dominant fraction of optically dark excitons. 
\end{abstract}

\date{\today}

\pacs{03.75.Lm,03.75.Mn,73.63.Hs,78.47.jd}

\maketitle

Massive bosonic particles realise a rich variety of collective quantum phenomena where their underlying fermionic structure is nevertheless hardly observed \cite{Leggett_2006,Stringari_Book_BEC}. For example,  Bose-Einstein condensation of atomic gases is generally understood by neglecting the atoms fermionic nature. Semiconductor excitons, i.e. Coulomb-bound electron-hole pairs, constitute a class of composite bosons which contrasts with this behaviour. Indeed, Combescot and co-workers have predicted that the fermionic structure of excitons leads to a multi-component condensate, with optically active and inactive parts that are coherently coupled through electron and/or hole exchanges between excitons \cite{Combescot_2007,Combescot_2012,Combescot_2015}. 

Widely studied GaAs quantum wells provide an interesting playground to demonstrate the predictions made by Combescot and co-workers, and then possibly conclude a fifty-year long quest for Bose-Einstein condensation of excitons \cite{Blatt_62,Keldysh_68,Zimmerman_08,Combescot_2016}. Indeed, in GaAs quantum wells lowest energy excitonic states exhibit a total "spin" ($\pm$1) or ($\pm$2). These states are then optically active and inactive respectively, dark states lying at the lowest energy. Neglecting exciton-exciton interactions, Bose-Einstein condensation then leads to a macroscopic occupation of dark states so that the condensate is completely inactive optically \cite{Combescot_2007}. Beyond a critical density, however, exciton-exciton interactions can dress the many-body ground-state. Fermion exchanges then become crucial because they can coherently convert opposite spin dark excitons into opposite spin bright ones \cite{Combescot_book}. Thus, a small bright component is possibly introduced coherently into the dark condensate \cite{Combescot_2012,Combescot_2015}. This results in a four-component many-body phase, which is grey, i.e. poorly active optically but possibly signalled by its weak photoluminescence coherent with the hidden dark part.

The dominantly dark nature of excitonic condensation manifests directly a high-temperature quantum phase transition. Indeed, in wide GaAs quantum wells the energy splitting between bright and dark states is of the order of $\mu$eV \cite{Blackwood}, i.e. small compared to the thermal energy ($\sim$2k$_\mathrm{B}$) at the condensation threshold \cite{Supplements}. As a result, a macroscopic population of dark excitons violates classical expectations. This point of view has long been overlooked by research of a condensate of bright excitons \cite{Zimmerman_08,Butov,Snoke,Gorbunov}, until recent works have instead pointed out experimentally the role played by dark states below a few Kelvin. These studies were realised with long-lived spatially indirect excitons \cite{Lozovik_76,Lozovik_97}, engineered by enforcing a spatial separation between electrons and holes, for instance by confining them in two adjacent GaAs quantum wells. Thus, a darkening of the photoluminescence has been reported below a few Kelvin \cite{Rapaport_2012}. Macroscopic spatial coherence of an anomalously dark gas has also been observed at sub-Kelvin temperatures \cite{Alloing_2014}.

Very recently, we have reported an important step towards unambiguous signatures for the dark state condensation of GaAs excitons \cite{Beian_2015}. Precisely, we have shown that indirect excitons can be confined in a 10 $\mu$m electrostatic trap and studied at controlled densities and temperatures, in a regime of vanishingly small inhomogeneous broadening.  This degree of control, never achieved before to the best of our knowledge, is necessary to evaluate the occupation of bright and dark states free from experimental uncertainties. Thus, we have shown unambiguously that the photoluminescence emission quenches below a critical temperature of about 1 Kelvin, when  $\sim$10$^4$ indirect excitons are trapped \cite{Ivanov_2010,Schindler_2008,Beian_2015}. The quenching was interpreted as the manifestation for the dark state condensation, however, the exact nature of the quantum phase remained inaccessible to these experiments relying on photoluminescence spectroscopy.

In this Letter, we report time and spatially resolved interferometry of the photoluminescence emitted by indirect excitons confined in a 10 $\mu$m trap, down to the regime of photoluminescence quenching. Below a critical temperature of about 1K, we demonstrate macroscopic spatial coherence and quantised vortices restricted to a small range of excitonic density, precisely in a dilute regime when 10$^4$- 2$\cdot$10$^4$ excitons are confined in the  trap. These superfluid signatures emerge for a population of bright excitons about 3 times smaller than the one of dark excitons. Our findings thus evidence quantitatively the theoretically predicted grey condensation of indirect excitons \cite{Combescot_2012}. This shows that bilayer GaAs heterostructures, either studied by photoluminescence \cite{Butov_trap,Butov_trap2,Holleitner_trap,Timofeev_trap,Cohen_2016,Stern_2014} or transport techniques \cite{Eisenstein_2012,Dietsche_2012,Ritchie_2008,Seamons_2009}, open a versatile platform to develop quantum control in semiconductors.

\begin{figure}[h!]\label{fig1}
\centerline{\includegraphics[width=.475\textwidth]{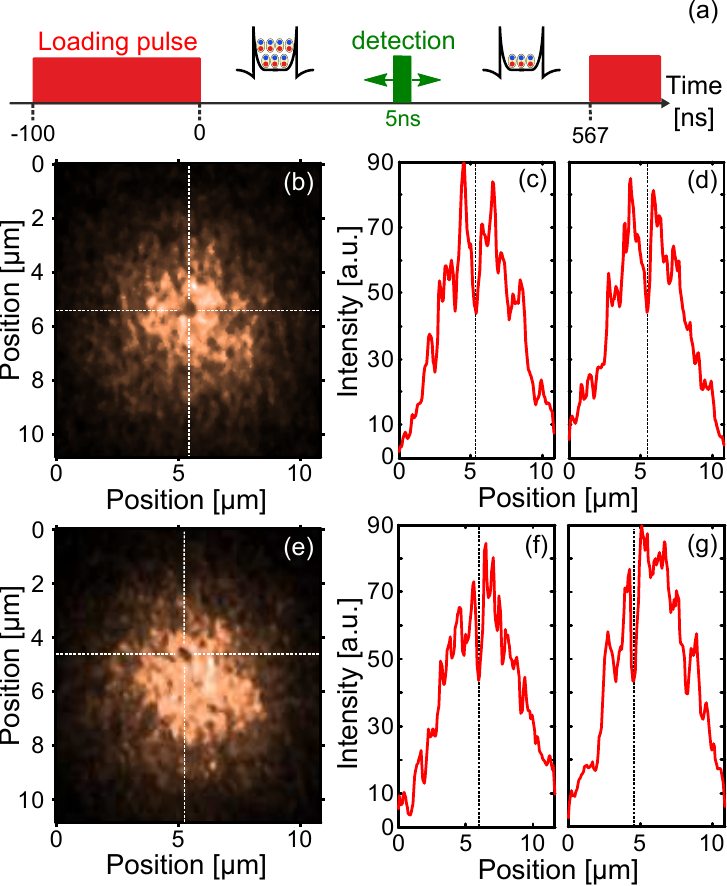}}
\caption{(a) A 100 ns long loading laser pulse injects indirect excitons in a 10 $\mu$m electrostatic trap. The reemitted photoluminescence is analysed in a 5 ns long detection window, at a variable delay to the end of the loading pulse, the sequence being repeated at 1.5 MHz for 10-20 seconds typical acquisition times. (b-e) Photoluminescence emitted from the trap,  at T$_\mathrm{b}$=330 mK and for a delay of 150 ns so that the number of trapped excitons is $\sim$2$\cdot$10$^4$.  Horizontal and vertical dashed lines highlight positions where we observe about 50 \% intensity loss along both horizontal and vertical axis. This is shown by the profiles in (c-d) and (f-g) for the images shown in (b) and (e) respectively. Measurements have all been acquired successively for identical experimental settings, the acquisition time being 10s.}
\end{figure}

As illustrated in Fig.1.a, our experiments rely on a 100 ns long laser pulse which loads indirect excitons in a shallow electrostatic trap. The latter is realised by controlling the electric field in the plane of two 8nm GaAs quantum wells, where photo-injected electrons and holes are confined (quantum wells being separated by a 4nm AlGaAs barrier -- see Supplementary Materials for more details). In the following, we emphasise the photoluminescence reemitted between 150 and 200 ns after extinction of the loading laser pulse. This delay range corresponds to about twice the indirect excitons optical lifetime \cite{Supplements}. During this time interval, the trapped gas is dilute and we estimate that the total number of excitons decreases from about 2$\cdot$10$^4$ to 10$^4$ . Thus, we detect spectroscopically a highly non-classical population of optically dark indirect excitons at sub-Kelvin bath temperatures \cite{Supplements,Beian_2015}. At the same time the photoluminescence emitted at the center of the trap is homogeneously broadened (see Fig.S1 of the Supplementary Materials). 

In Figure 1.b we show the spatial profile of the photoluminescence emitted when $\sim$ 2$\cdot$10$^4$ excitons are trapped at a bath temperature T$_\mathrm{b}$=330 mK. We strikingly note a very inhomogeneous intensity distribution, a dark spot being identified at the centre of the image, i.e. at the minimum of the trapping potential where the photoluminescence intensity is nevertheless the largest. At the centre of the dark spot we observe 50$\%$ loss of intensity (Fig.1.c-d) corresponding to a 2-fold decrease of the population of bright excitons. This variation marks a deviation of $\sim$5$\sigma$ of the photoluminescence signal which is not interpretable in terms of intensity fluctuations. 

In our experiments, the unambiguous detection of dark spots, as in Figure 1.b, requires precise experimental settings. It is mostly achieved around the center of the trap, at sub-Kelvin bath temperatures and for less than about 4$\cdot$10$^4$ confined excitons, that is later than 120 ns after extinction of the loading pulse. Experimentally, a statistically unambiguous detection of dark spots resumes to a tradeoff between the signal to noise ratio and the number of individual realisations that we average, that is the acquisition time. The latter can not exceed about 10 seconds, because at T$_\mathrm{b}$=330 mK  dark spots emerge at uncorrelated positions during unchanged experimental settings. This behaviour is signalled by comparing the emission profiles shown in Fig. 1.b and 1.e. Both were recorded successively and in the same conditions, nevertheless they exhibit intensity losses localised at distinct positions in the central region of the trap. 

We interpret the dark spots in the photoluminescence as a direct manifestation for the disorder of our electrostatic confinement. In Ref.\cite{Beian_2015} we have already highlighted that the trapping potential fluctuates during our experiments. The level of electrostatic disorder is such that it leads to stochastic variations of the photoluminescence spectral width, from $\sim$300 $\mu$eV to $\sim$1 meV and within a timescale of a few seconds at T$_\mathrm{b}$=330 mK. However, the electrostatic disorder can be turned into an advantage to signal quantum fingerprints for the regime of photoluminescence quenching \cite{Beian_2015}. Indeed, defects of the confining potential are energetically favourable positions to localise quantised vortices and thus reveal a superfluid behaviour. Vortices could then remain pinned in the trapping potential, the only situation to actually detect them by our experiments which rely on averaging $\sim$ 10$^7$ single-shot images during 10 seconds.

\begin{figure}
\centerline{\includegraphics[width=.5\textwidth]{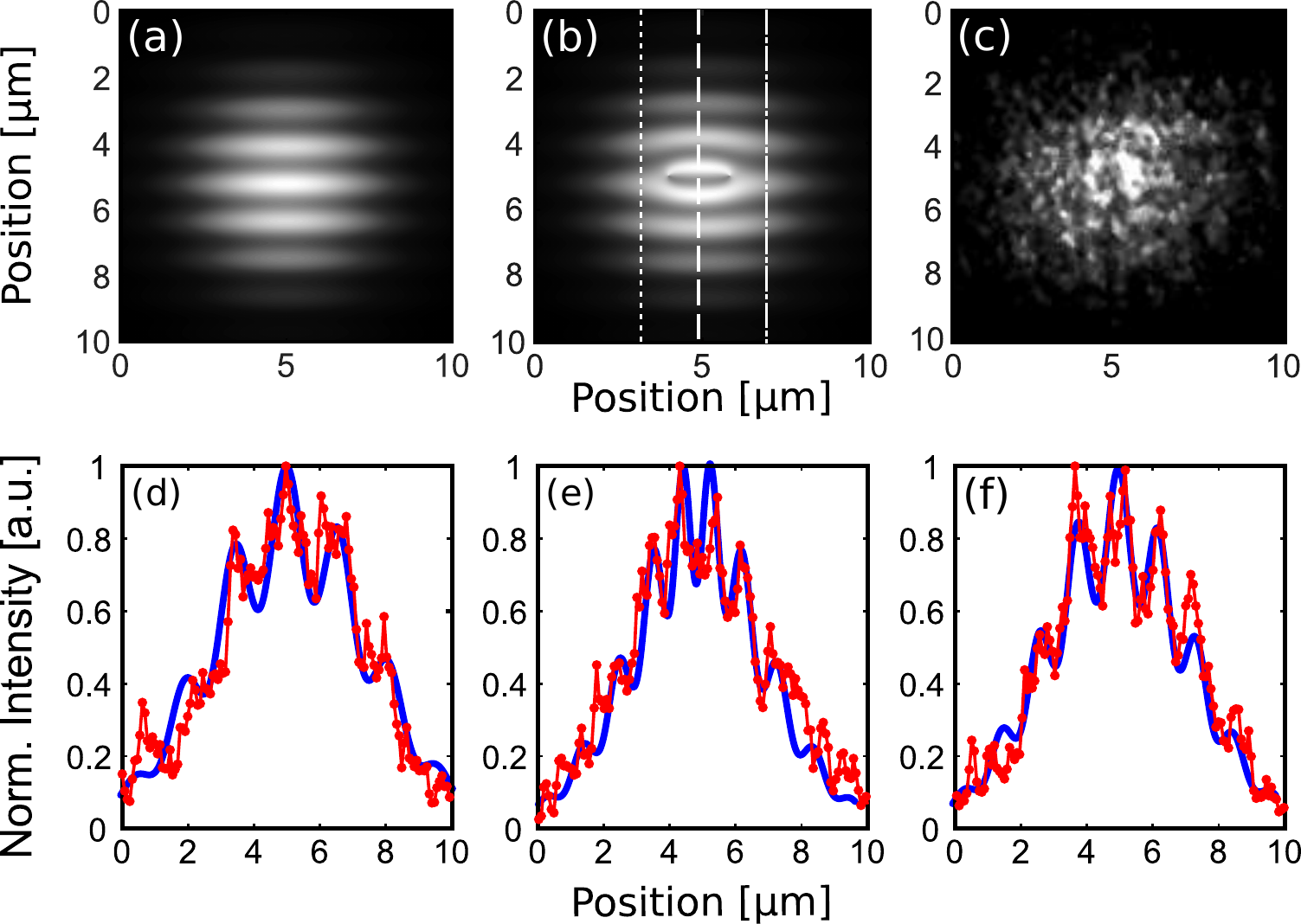}}
\caption{Simulation of the interference pattern for a condensate with complete long-range order in the trap (a), and for a condensate constraining one quantised vortex at the center of the trap (b). Two phase singularities are observed in the latter case, on each side of a ring-shaped interference fringe. (c) Interference pattern measured when $\sim$ 2$\cdot$10$^4$ excitons are confined in the trap at T$_\mathrm{b}$=330 mK. These experiments were realised in the same conditions as for the measurements shown in Fig. 1.b. (d-f) Red points show the interference profiles measured, as highlighted in (b), at the centre of the ring (e), on its left (d) and right (f). The solid blue lines display the patterns simulated by modulating the profile of the photoluminescence intensity with an interference visibility equal to 23$\%$, the interference contrast possibly varying from $\sim$12 $\%$ to 45 $\%$ in our studies.}
\end{figure}

To asses whether dark spots detected at the center of the trap can manifest quantised vortices pinned by electrostatic disorder, we analysed the spatial coherence of the photoluminescence with a Mach-Zehnder interferometer. The interferometer is stabilised with a vanishing path length difference between its two arms, one of which horizontally displaces its output   by 2 $\mu$m compared to the other arm \cite{Alloing_2014}, i.e. by ten times the thermal wavelength of excitons at our lowest bath temperature. Due to a vertical tilt angle deliberately introduced between the two arms, a condensate with complete long-range order leads to horizontally aligned interference fringes (Fig.2.a). A vortex pinned around the center of such condensate then appears through the inclusion of a "ring" in the central bright fringe, as shown in Fig. 2.b. This pattern is understood by noting that on each side of the ring the vortex and its shifted image interfere with 2 $\mu$m distant regions where the phase is well defined. Fork-like dislocations are thus created at these two locations since the phase of the wavefunction winds by 2$\pi$ around the core of a vortex \cite{Stringari_Book_BEC}. The superposition of the two mirrored and shifted forks leads then to the "ring" shown in Fig. 2.b, making this interference pattern topologically recognisable. 

Figure 2.c shows an interference pattern measured in the same conditions as for the experiments of Fig. 1.b. Remarkably, this observation agrees quantitatively with the simulation for a condensate having one quantised vortex pinned at the center of the trap. This is shown in  Fig. 2.d-f where the interference profiles taken at the centre of the ring (e) and on its left and right, (d) and (f) respectively, are reproduced by modulating the photoluminescence intensity profile with 23$\%$ interference visibility. The contrast providing the fraction of bright excitons in the superfluid phase \cite{Glauber_99}, we deduce that about one third of bright excitons evolve in a quantum condensed state for these experiments. Let us then stress that the results shown in Fig. 2.c are obtained by post-selecting a particular realisation out of successive acquisitions, measured all under the same conditions. Such a post-selection is necessary because our studies suffer from electrostatic fluctuations. In fact, it is only for a particular confinement landscape that an individual vortex is possibly revealed, as in Fig. 2.c. The electrostatic trapping potential has to be sufficiently regular for a superfluid to possibly form, and exhibit a defect capable to localise a single vortex around the center of trap, the position of this defect being stable all along the measurement time.

\begin{figure}\label{fig3}
\centerline{\includegraphics[width=.5\textwidth]{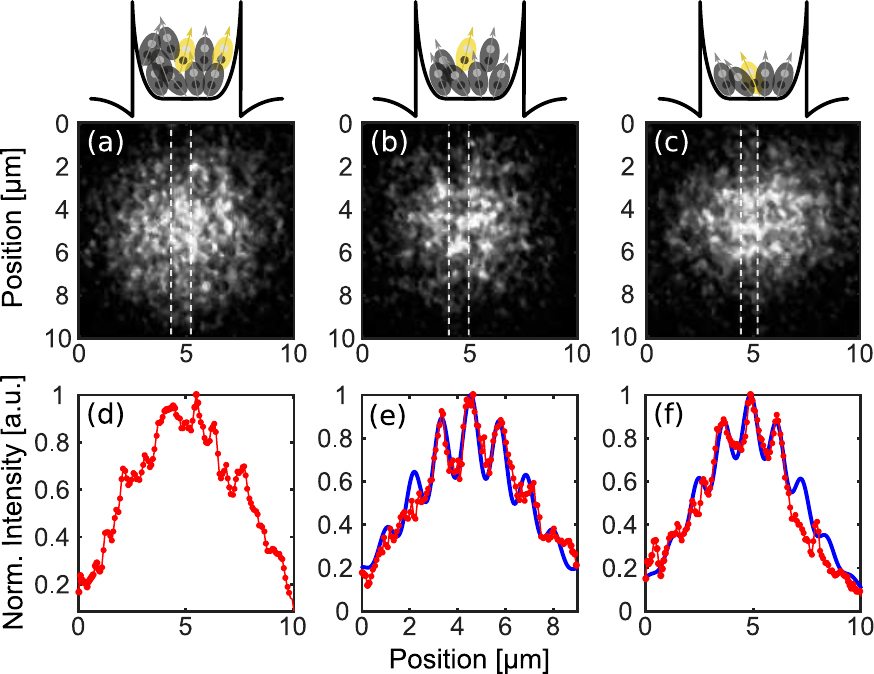}}
\caption{(a-c) Interference patterns measured for a decreasing exciton density in the trap at T$_\mathrm{b}$=330 mK, 120 (a), 150 (b) and 200 ns (c) after extinction of the loading laser pulse. We estimate that the total number of excitons is about  4$\cdot$10$^4$, 2$\cdot$10$^4$ and 10$^4$ respectively, the drawings on top illustrating the filling of the trap. The panels (d) to (f) show the corresponding interference profiles evaluated at the center of the trap (between the dashed lines). While in (d) our experiments do not reveal any interference, in (e) and (f) the interference visibility is 25 and 18 $\%$ respectively. Red points show experimental results and the blue lines the simulation obtained by modulating the intensity profile with the aforementioned visibilities.}
\end{figure}

Without varying experimental conditions we also studied the evolution of quantum coherence while the exciton density varies in the trap. This is directly achieved by changing at the detection the delay to the end of the loading laser pulse (Fig.1.a). To reach conclusions that are not  limited by potential fluctuations during our measurements, we successively recorded a set of 20 interference patterns, every 10 ns after the loading pulse. Fig. 3.a shows that for delays shorter than 150 ns, i.e. when the trap confines more than about 2$\cdot$10$^4$ indirect excitons, interference fringes are not resolved in the photoluminescence. By contrast, from 150 to 200 ns after optical loading, i.e., when the population of excitons in the trap decreases from 2$\cdot$10$^4$ to 10$^4$, Fig.3.b-c shows that bright excitons exhibit macroscopic spatial coherence: interference fringes are clearly resolved in patterns that cover the center of the trap, i.e. an approximately 5x5 $\mu m^2$ region. At longer delays ($\gtrsim$ 200 ns), however, interference fringes are not detected clearly in our experiments.

The absence of interference pattern when the trap confines less than about 10$^4$ excitons is not very surprising. Indeed, in this regime repulsive interactions between excitons yield a low mean-field energy, of the order of potential fluctuations ($\sim$ 500 $\mu$eV \cite{Beian_2015}). The trapped gas is then probably too dilute to establish long-range coherence by screening electrostatic disorder \cite{Berman_2004}. On the other hand, it is more surprising that quantum coherence is not observed beyond a maximum of about 2$\cdot$10$^4$ particles in the trap. Yet, this limit lies well in the dilute regime which excludes the role of exciton ionisation. However, excitons may already suffer from a too large deviation to ideal bosons beyond this range of density \cite{Monique_2001}. Also, one can not exclude that beyond 2$\cdot$10$^4$ particles in the trap the strong dipolar  interactions between excitons already lead to correlations which challenge the emergence of a collective quantum phase.

Last, we studied the dependence of the interference contrast as a function of the bath temperature. Let us restrict ourselves to the relevant range of delays to the loading laser pulse (150 to 200 ns). For the shortest delay, i.e. for $\sim$2$\cdot$10$^4$ indirect excitons in the trap, Fig. 4.a shows that the photoluminescence exhibits long range order at the center of the trap, up to a critical temperature T$_\mathrm{c}\approx$ 1.3K. The interference visibility, i.e., the fraction of bright excitons contributing to the superfluid, follows well the theoretical scaling proportional to $1-($T$_\mathrm{b}/$T$_\mathrm{c})^2$ for two-dimensional particles in a trap \cite{Bloch_2008}. Furthermore, Fig.4.b shows that T$_\mathrm{c}\sim$1K when the density is decreased by around two-fold, i.e. at a delay of 200 ns after the loading pulse. This decrease of T$_\mathrm{c}$ is expected \cite{Bloch_2008}, however, quantitative conclusions are difficult to raise since our experiments are limited by the weak photoluminescence intensity. As underlined in Fig. 4, our measurements suffer from a signal-to-noise ratio of less than 10 which leads to a minimum threshold for our interferometric detection of about 12$\%$. Experiments displaying no evidence of spatial interference are then assigned 12$\%$ visibility.

\begin{figure}\label{fig4}
\centerline{\includegraphics[width=.45\textwidth]{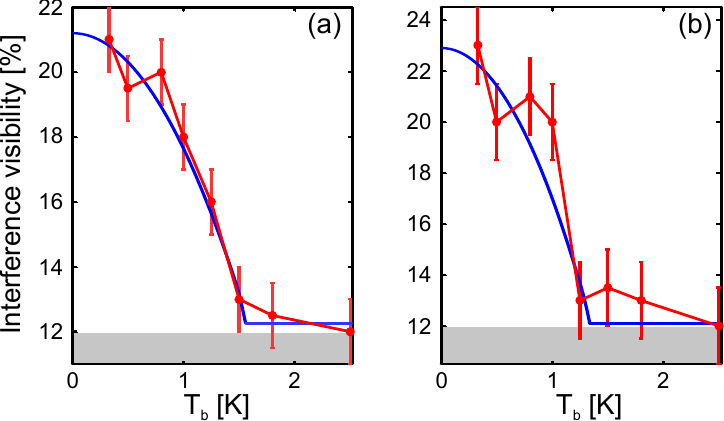}}
\caption{Interference contrast measured at the centre trap as a function of the bath temperature T$_\mathrm{b}$. In (a) we show the visibilities measured in the regime where 2$\cdot$10$^4$ excitons occupy the trap, while the number of excitons is reduced to 10$^4$ for the measurements shown in (b). Solid red lines show the theoretically expected 1$-$(T/T$_\mathrm{c}$)$^2$ scaling of the condensate fraction, with  T$_\mathrm{c}\sim$ 1.3K and 1K for (a) and (b) respectively. In (a)-(b) the grey region marks the sensibility of our interferometric detection, i.e. the level fixed by the signal-to-noise ratio at the detection.}
\end{figure}

Although quantum signatures are detected in the photoluminescence emitted by bright exciton states, crude estimations show that their occupation is too small to allow for a bright condensate independent from the underlying dominant population of optically dark excitons. Indeed, out of $\sim$10$^4$ excitons confined at T$_\mathrm{b}$=330 mK, about 3/4 populate dark states \cite{Supplements}. By only considering the remaining fraction of bright excitons, the critical temperature for quantum degeneracy would be less than $\sim$300 mK \cite{Supplements}. A fragmented condensate of bright excitons would then contradict our experiments which, as shown in Fig. 4, reveal quantum coherence up to 1.3K, as expected for a few 10$^4$ excitons in the trap. Considering limiting factors, such as the strength of electrostatic disorder, it is actually excluded that such a low density of bright indirect excitons possibly condenses alone \cite{Berman_2004}. This leads us to conclude that dark and bright states are coherently coupled in our experiments, leading to the theoretically predicted four-component superfluid of indirect excitons \cite{Combescot_2012}.

Finally, let us note that experiments with cold atomic gases have recently explored the superfluid quantum phase transition, by cooling a Bose gas at a variable rate. It was hence verified that the size of superfluid domains formed at the critical point decreases with the quenching rate \cite{Weiler_2008,Navon_2015}, as prescripted by the Kibble-Zurek mechanism \cite{Zurek_96}. Here, we had to follow the opposite approach, because the bath temperature can be kept constant while the exciton density necessarily decreases slowly, due to radiative recombination. Thus, we observe that an initially dense gas, showing no evidence of long-range coherence, abruptly becomes superfluid below a critical density of a few 10$^{10}$ cm$^{-2}$ at sub-Kelvin temperatures. In this regime, quantum signatures are resolved in the coherent photoluminescence radiated by the four-component and mostly dark condensate of excitons. Interestingly, this behaviour is restricted tor a narrow range of densities only.

\textbf{Acknowledgements:} The authors are grateful to Monique Combescot and Roland Combescot for their continuous support of this work and for many enlightening discussions. We would also like to thank Tristan Cren for stimulating discussions and Maciej Lewenstein for a critical reading of the manuscript. Our work has been financially supported by the projects INDEX (EU-FP7-ITN), XBEC (EU-FP7-CIG) and by OBELIX from the french Agency for Research (ANR-15-CE30-0020). Correspondence and requests shall be sent to F.D. (francois$\_$dubin@icloud.com).


\begin{center}
\Large{\textbf{Supplementary Informations:}}
\end{center}

\section{Sample structure and details of the electrostatic trap}

The sample studied here is identical to the one probed in Ref. \cite{Beian_2015}. It mainly consists of two 8 nm wide GaAs quantum wells which are separated by a 4 nm AlGaAs barrier layer. The two coupled quantum wells (CQWs) are positioned 150 nm above the n-doped GaAs layer that acts as bottom electrode of the field-effect device embedding the CQWs. To realise a 10 $\mu$m wide electrostatic trap, we use a set of 2 semi-transparent and metallic electrodes deposited on the surface of the field-effect device, i.e., 900 nm above the CQWs. In this geometry, the components of the electric-field applied by the gate electrodes are minimised in the plane of the CQWs which prevents undesired exciton dissociation.

\vspace{.5cm}
\centerline{\includegraphics[width=.45\textwidth]{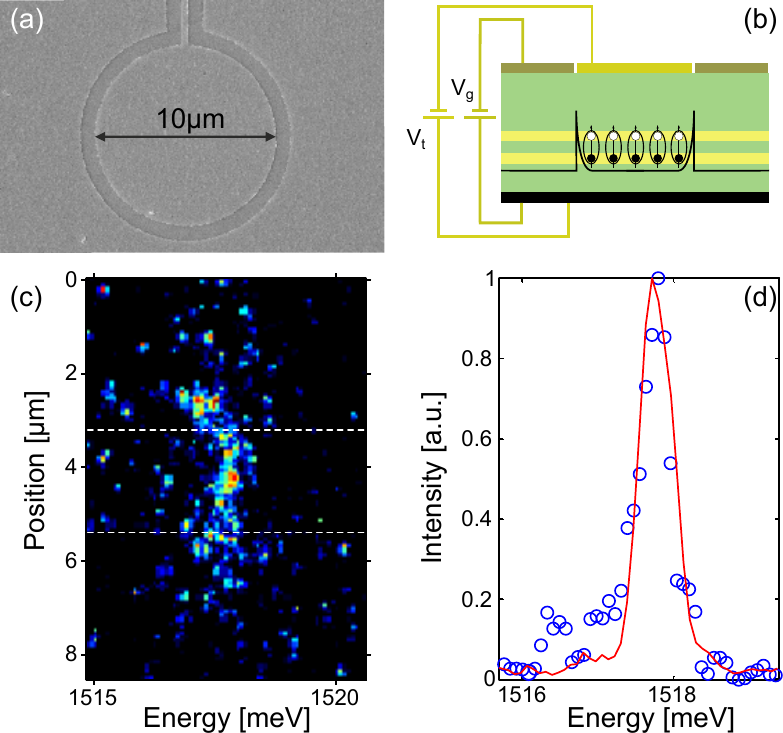}}
\vspace{.1cm}
\textbf{Fig. S1:} \textit{(a) Electron microscope image of the two surface electrodes controlling our electrostatic trap. In our experiments these electrodes are biased at about -4.8V; so, we realise a hollow-trap, i.e., a shallow trap characterised by the barrier due to the potential rectification at the 200 nm gap between the surface electrodes. (b) Side-cut of our sample structure showing the two coupled quantum wells (in yellow) embedded in the field-effect device. Electrons and holes confined in each layer bind by Coulomb attraction in this way forming spatially indirect excitons characterised by electric dipoles aligned with the externally applied electric field. In (b), we also sketch the profile of the hollow-trap by the thin solid black line. (c) Spatially resolved photoluminescence spectrum recorded at T$_\mathrm{b}$=330 mK for the same experimental conditions as for the measurements shown in Fig. 1. In (d), we show the corresponding spectral-profile, evaluated between the dashed lines shown in (c), together with our spectral resolution measured with a Hg line (red).}\\

The general structure of our electrostatic trap is illustrated in Fig. S1. The trap is engineered by a circular 10 $\mu$m wide central electrode separated by a 200 nm gap from its outer guard gate. Far from the edges of the electrodes, e.g., at the centre of the trap or under the guard, we have verified that the electric field amplitude perpendicular to the CQWs is accurately controlled by the static potentials we externally applied to each electrode. By imposing a bias onto the trap gate larger than the one applied onto the guard gate, deep electrostatic traps are formed for indirect excitons (IXs) \cite{High_2009,Holleitner_2013}. Indeed, Fig. S1 shows that IXs are characterised by their intrinsic electric dipole aligned perpendicular to the CQWs. They thus behave as high-field seekers, that is, they are attracted towards the regions of the CQWs where the perpendicular electric field is the largest. 

Hard walls to confine indirect excitons in a shallow trapping potential can also be engineered with our device. Indeed, electrostatic barriers form spontaneously under the gap between our surface electrodes. We have verified that the barriers height amounts to at least 10 meV for IXs, even if the trap and the guard gate are polarised at the same potential. Under the trap gate the electrostatic confinement is then shallow and regular, that is why we have decided to apply the same bias to the trap and guard gates. We found a particular value of $\sim$-4.8V for which the dark current of the device was vanishingly small while the steady-state photo-current would not exceed 100 pA for our measurements at T$_\mathrm{b}$=330 mK. Fine experimental settings were motivated by the search for the spectrally narrowest photoluminescence. In particular, this implied that we aimed at the most stable conditions with the smallest amount of electrostatic fluctuations during our measurement sequence (15-30 seconds for a single acquisition). This approach brought us to regimes where the photoluminescence spectra are limited by our spectral resolution across the center of the trap when we observe superfluid signatures. This is shown in Fig. S1.c-d that displays a spatially resolved photoluminescence emission recorded under the same conditions as for the measurements shown in Fig. 1.b of the main text. 

\section{Experimental procedure}

As in previous works, to optically inject  electrons and holes in the CQWs we used a pulsed laser excitation tuned at resonance with the absorption of direct excitons of the two quantum wells. As shown in Fig. S2, each pulse loads the two quantum wells with both electrons and holes. The electric field applied perpendicular to the heterostructure favours carriers tunnel towards their respective minimum energy states which lie in different quantum wells. As a result, spatially indirect excitons are formed by the Coulomb attraction between  electrons in one layer and holes in the other layer. Indirect excitons are formed in few tens of ns and constitute the majority carriers already at the end of the laser excitation. In the case of our experiments probing a finite size electrostatic trap, we have shaped the spatial profile of the laser excitation such that it homogeneously covers the bottom of the trap which is about 5 $\mu$m wide. At the same time we ensured that the illumination outside the trap was negligible.

\centerline{\includegraphics[width=.45\textwidth]{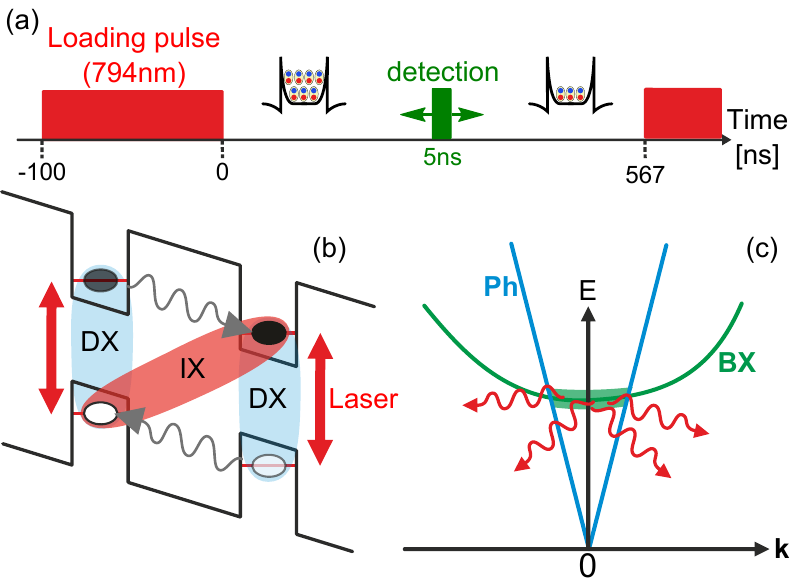}}
\textbf{Fig. S2:} \textit{(a) Our experiments rely on a 100 ns long loading pulse while the exciton dynamics is monitored in a 5 ns long time window which follows each laser pulse at a variable delay. (b) Sketch of the optical injection of indirect excitons IXs, through the resonant excitation of the direct exciton (DX) absorption of each quantum well. IXs are created once optically injected electronic carriers have tunnelled towards minimum energy states (wavy grey lines). (c) The reemitted photoluminescence (wavy red arrows) is only due to the radiative recombination of lowest energy (\textbf{k}$\sim$0) bright excitons (BX), i.e., lying at an energy smaller than the intersection between bright excitonic and photonic bands, BX and Ph respectively, occurring at E$\sim$1.5 k$_\mathrm{B}$.}\\

\section{Quantum darkening and density of cold indirect excitons}

Figure S2 shows that our experiments rely on a 100 ns long laser excitation which is repeated at a rate of 1.5 MHz. All measurements have been performed with a mean optical power equal to 700 nW, the incident laser power being actively stabilised in our studies. The photoluminescence emitted from the trap after the optical loading phase was directed towards an imaging spectrometer that allows us to study the spatial profile of the emission either in real or frequency space. As detailed in Ref. \cite{Beian_2015}, in the latter case we monitor both the energy and integrated intensity of the photoluminescence in order to estimate the fraction of bright and dark indirect excitons in the trap. The dynamics of the photoluminescence energy E$_\mathrm{X}$ reflects the variation of the total exciton density n$_\mathrm{X}$, i.e., including both bright and dark excitons. Indeed, indirect excitons experience repulsive dipolar interactions in the dilute regime such that E$_\mathrm{X}$ scales as $u_0$n$_\mathrm{X}$ at first order \cite{Ivanov_2010,Schindler_2008}, $u_0\sim$1meV for the density n$_\mathrm{X}$ in  10$^{10}$ cm$^{-2}$ units at which the trap confines $\sim$10$^4$ excitons. On the other hand, the sole fraction of bright excitons at lowest energy is directly given by the integrated intensity of the photoluminescence I$_\mathrm{X}$. Indeed, only bright indirect excitons with a kinetic energy lower than about 1.5 K contribute to the photoluminescence. This region of the excitonic band is usually referred to as the light cone and reduced to excitons with a vanishing in-plane momentum \textbf{k} (see Fig.S2). \\

\centerline{\includegraphics[width=.5\textwidth]{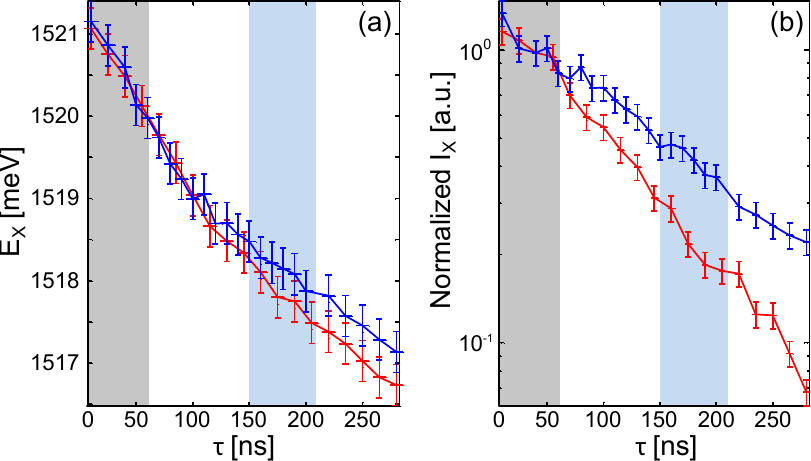}}
\textbf{Fig. S3:} \textit{(a) Dynamics of the photoluminescence energy E$_\mathrm{X}$, measured after the loading laser pulse at the centre of the trap at T$_\mathrm{b}$=330 mK and 2.5 K, red and blue respectively. (b) Integrated intensity I$_\mathrm{X}$ for the same experiments. In (a-b) the initial grey region underlines a transient regime where we can not exclude that the trapped gas is not fully thermalised and also subject to a transient photocurrent. The light-blue regions underline the regime where we observe superfluid signatures.}\\

Figure S3 shows the dynamics of both I$_\mathrm{X}$ and E$_\mathrm{X}$ starting from 5 ns after the extinction of the loading laser pulse. In Fig. S3, we compare two limiting cases, namely a thermal gas of excitons realised at a bath temperature T$_\mathrm{b}$=2.5 K, and a quantum gas of excitons realised at  T$_\mathrm{b}$=330 mK. These two measurements are performed under the same experimental conditions, that is, the same laser mean excitation-power and the same voltages applied onto the gate electrodes. However, these measurements differ by the anomalous darkening in the dynamics of I$_\mathrm{X}$ which is observed at the lowest temperature. Indeed, Fig. S3.b shows that the integrated intensity decays 50$\%$ faster at T$_\mathrm{b}$=330 mK than at 2.5 K. Precisely, after a transient regime lasting about 50 ns after the laser excitation, I$_\mathrm{X}$ has a characteristic decay time of $\sim$ 100 and 160 ns at 330 mK and 2.5 K respectively.  Thus, 150-200 ns after the laser pulse I$_\mathrm{X}$ is 2 times weaker at T$_\mathrm{b}$=330 mK revealing a strong depletion of coldest bright indirect excitons. By contrast, Fig. S3.a shows that at T$_\mathrm{b}$=330 mK and 2.5 K the photoluminescence energy E$_\mathrm{X}$ follows close dynamics which signals that the total density in the trap varies weakly between these two measurements. As detailed in Ref \cite{Beian_2015}, these combined observations reveal without ambiguity that a dominant fraction of IXs populates optically dark states at T$_\mathrm{b}$=330 mK. The large imbalance between bright and dark states occupation is highly non-classical (75$\%$ in dark states) because the energy splitting between these states reduces to a few $\mu$eV in our heterostructure \cite{Blackwood}, that is about 10-fold less than the thermal energy at our lowest bath temperature.

\section{Interferometric measurements}

To quantify the first order spatial coherence of bright indirect excitons we analyzed the photoluminescence emitted from the trap with a Mach-Zehnder interferometer. The  photoluminescence was splitted between the arms 1 and 2 of the interferometer, and a vertical tilt angle $\alpha$ was deliberately introduced between the outputs of the two arms. Hence, interference fringes are aligned horizontally, $\alpha$ being set such that the interference period is $\approx$ 1.5 $\mu$m. From the spatial auto-correlation, the outputs produced by the two arms are laterally shifted by $\delta_x$= 2$\mu$m while the path length difference is stabilized close to zero. This allows us to derive the degree of spatial coherence of bright IXs which thermal de Broglie wavelength is bound to less than 300 nm for 10$^4$ excitons in the trap at T$_\mathrm{b}$=330 mK. Thus, a thermal gas of excitons leads to a vanishing interference contrast, as expected and verified at  T$_\mathrm{b}\sim$2K in Fig. 4 of the main text.

The output of our interferometer, I$_{12}$, can be modelled as 
\begin{equation}
I_{12}(\textbf{r};\delta_x)=\langle|\psi_0(\textbf{r},t)+e^{i(q_\alpha
y+\phi)}\psi_0(\textbf{r}+\delta_x,t)|^2\rangle_t \nonumber
\end{equation}
where $\psi_0$(\textbf{r}) is the photoluminescence field which reflects the bright excitons wave function, $\langle..\rangle_t$ denotes the time averaging, \textbf{r}=(x,y) is the coordinate in the plane of the quantum well and
$q_\alpha$=2$\pi\lambda^{-1}$sin$(\alpha)$ with $\lambda$ being the emission wavelength. If we denote the
output of the two arms by I$_{1}$ and I$_{2}$ respectively, we directly deduce that
I$_\mathrm{int}$=(I$_{12}$-I$_{1}$-I$_2$)/2$\sqrt{I_{1}I_2}$ reveals the first order coherence function of indirect excitons, defined as
\begin{equation}
 g^{(1)}(\textbf{r};\delta_x)=\frac{\langle\psi^*_0(\textbf{r},t)\psi_0(\textbf{r}+\delta_x,t)\rangle_t}{
(\langle|\psi_0(\textbf{r}
,t)|^2\rangle_t\langle|\psi_0(\textbf { r }
+\delta_x,t)|^2\rangle_t)^{1/2}}.\nonumber
\end{equation}
Indeed, I$_\mathrm{int}$(\textbf{r};$\delta_x$)=cos($q_\alpha
y+\phi+\phi_\mathrm{\textbf{r}}$)$|g^{(1)}(\textbf{r};\delta_x)|$ where
$\phi_\mathrm{\textbf{r}}$ is the phase of $g^{(1)}$. Thus, we recover that interference fringes have a visibility controlled by the degree 
of spatial coherence of bright excitons. On the other hand, the position of the interference fringes reveals the argument of the $g^{(1)}$-function, i.e., the phase difference between the interfering wave functions. 

\vspace{.3cm}
\centerline{\includegraphics[width=.5\textwidth]{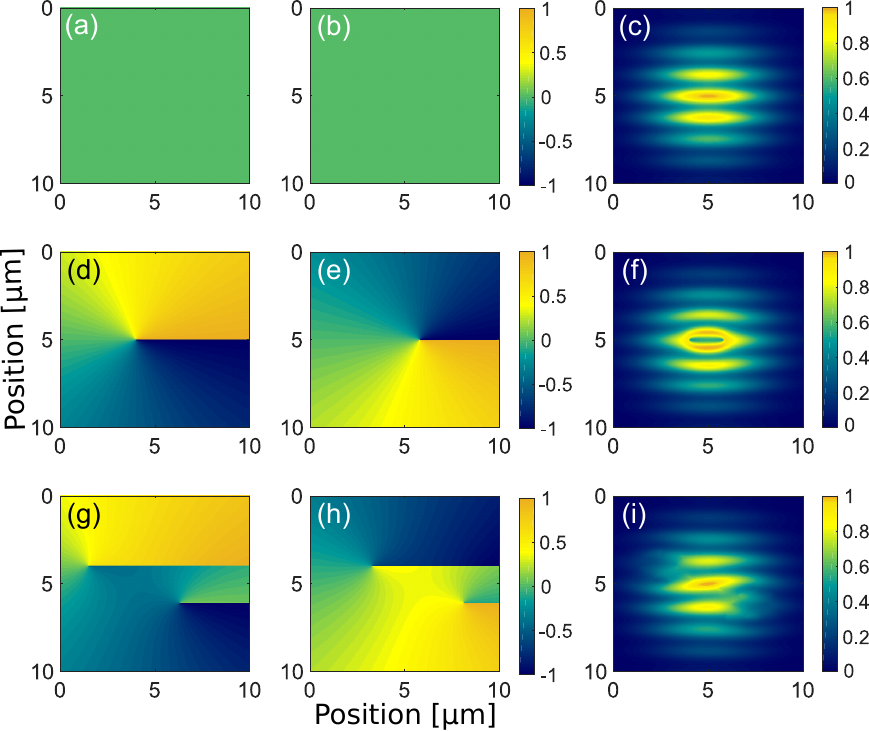}}
\textbf{Fig. S4:} \textit{ (a-b) Phase distribution for the photoluminescence fields entering the g$^{(1)}$-function, computed for a condensate with complete long-range order. The panel (c) shows the resulting interference profile. (d-e) Phase profiles of interfering fields for a condensate  constraining one vortex. The panel (f) shows the resulting interference pattern. (g-h) Phase profiles for an average of 2 vortices which explore a few microns extended regions. The position of each vortex is given by  the locations of phase singularities in (g-h). In (i) we note that the diffusion of vortices blurs and also bends interference fringes.}

\vspace{.2cm}
Interference patterns are directly simulated by modulating the spatial profile of the photoluminescence emission, with the above expression for I$_\mathrm{int}$. Starting with the simplest case, Figure S4.c  considers a condensate of bright excitons with complete long-range spatial coherence. As a result, the phase of $\psi_0$ is uniform and the output interference pattern consists of horizontally aligned fringes. Figure S4.f shows a second  scenario where one quantised vortex is constrained at the center of such a condensate. As mentioned in the main text, the phase of $\psi_0$ then winds by 2$\pi$ around the core of the vortex. The interference between such singularity and its shifted mirror image then leads to the "ring" fringe shown in Fig. S4.f. Finally, in Fig. S4.g-i, we study a situation which probably matches best the conditions under which our experiments are performed. Indeed, we consider two vortices  that can diffuse over a few microns in the condensate. Fig. S4.i shows, as expected, that moving vortices tend to blur the interference signal, in a fashion which resembles the measurements shown in Fig. 3.b. \\

-- \textbf{Analysis of interference patterns}
\vspace{.1cm}
\newline

In the main text we only report bare interference measurements, i.e. without applying any correction to the experimental data. These latter are simulated by fitting the two-dimensional profile of the photoluminescence intensity $\langle |\psi_0(\textbf{r})|^2\rangle_t$ which is then modulated according to the above expression for I$_\mathrm{int}$. The interference visibility $|g^{(1)}|$ is then extracted by reproducing the profiles taken along the vertical axis perpendicular to the fringe direction. 

Let us finally note that we have set the Mach-Zehnder interferometer for $\delta_x$= 2 $\mu$m, first because the condensate extends over a region of about 5x5 $\mu m^2$. Also, this value lies well above our optical resolution ($\sim$ 1 $\mu$m) which is limited by mechanical vibrations in our cryostat. Thus, for $\delta_x$= 2 $\mu$m interference fringes form accros a region which is sufficiently large to be quantitatively analysed. For larger values of $\delta_x$, however, the overlap between interfering regions is reduced further which hardens a quantitative discussion of quantum coherence. 

\section{Estimations for Bose-Einstein condensation of spatially indirect excitons}

In GaAs coupled quantum wells, indirect excitons are characterised by an effective mass M$_\mathrm{X}$ that is of the order of 0.2m$_0$, m$_0$ being the electron mass. In addition, indirect excitons have a Bohr radius a$_\mathrm{X}$ estimated to be $\sim$ 20 nm for our heterostructure and experimental settings. The regime in which we have typically 10$^4$ excitons in a 10 $\mu$m wide trap is dilute since n$_\mathrm{X}$a$_\mathrm{X}^2\lesssim$ 0.2. The inter-exciton mean separation is then large compared to the Bohr radius.

By neglecting exciton-exciton interactions and fermion exchanges, we can get a crude estimate of the temperature T$_\mathrm{0}$ for quantum degeneracy of a homogeneous gas of indirect excitons. For that purpose, we equalise the thermal de Broglie wavelength to the inter-particle distance. This leads to k$_\mathrm{B}$T$_\mathrm{0}$=n$_\mathrm{X}$($2\pi\hbar^2$)/($gM_\mathrm{X}$), where $g$ denotes the degeneracy of the considered excitonic states. If we first consider 10$^4$ dark excitons, i.e. n$_\mathrm{X}$=10$^{10}$ cm$^{-2}$, for which $g$=2 and thereby T$_\mathrm{0}\sim$ 1.3K. By contrast, if we consider the fraction of bright excitons, which is damped to at most 3$\cdot$10$^3$ in our measurements, the critical temperature drops by at least three-fold and then lies in the range of our lowest bath temperature. Our measurements being notably limited by electrostatic fluctuations of the trapping potential, with an amplitude of about 500 $\mu$eV, it is excluded \cite{Berman_2004} that bright excitons can  condense independently from the underlying condensate of dark excitons.


\begin{thebibliography}{99}
 
 \bibitem{Leggett_2006} A.J. Leggett \textit{Quantum Liquids} (Oxford. Univ.
Press, 2006)  

\bibitem{Stringari_Book_BEC} \textit{``Bose-Einstein
Condensation''}, L.P. Pitaevskii, S. Stringari (Oxford university Press, 2003)

\bibitem{Combescot_2007} M. Combescot, O. Betbeder-Matibet, R. Combescot,
Phys. Rev. Lett. \textbf{99}, 176403 (2007)

\bibitem{Combescot_2012} R. Combescot and M. Combescot, Phys. Rev. Lett. \textbf{109}, 026401 (2012)

\bibitem{Combescot_2015}  M. Combescot, R. Combescot, M. Alloing, F. Dubin, Phys. Rev. Lett. \textbf{114}, 090401 (2015)

\bibitem{Blatt_62} J. M. Blatt et al., Phys. Rev. \textbf{126}, 1691 (1962)

\bibitem{Keldysh_68} L.V. Keldysh and A.N. Kozlov, Sov. Phys. JETP \textbf{27},
521 (1968)

\bibitem{Zimmerman_08} R. Zimmerman \textit{"Bose-Einstein condensation of excitons: promise and disapointment"} (Oxford. Univ. Press., Eds. A. L. Ivanov and S. G. Tikhodeev, 2008) 

\bibitem{Combescot_2016} M. Combescot, R. Combescot, F. Dubin, Rep. Prog. Phys. (in press)

\bibitem{Combescot_book} M. Combescot and S.Y. Shiau "Excitons and Cooper Pairs: Two Composite Bosons in Many-Body Physics" (Oxford. Univ. Press, 2016)

\bibitem{Blackwood} E. Blackwood, M. Snelling, R. Harley, S. Andrews, C. Foxon, Phys. Rev. B \textbf{50}, 14246 (1994)

\bibitem{Supplements} see Supplementary Materials

\bibitem{Butov} L. V. Butov, JETP \textbf{122}, 434 (2016) and references therein

\bibitem{Snoke} D Snoke, Physica Status Solidi (b) \textbf{238}, 389 (2003) and references therein

\bibitem{Gorbunov} A. V. Gorbunov and V. B. Timofeev, J. Low Temp. Phys. \textbf{42}, 340 (2016) 

\bibitem{Lozovik_76} Y.E. Lozovik, V.I. Yudson, Zh. Eksp. Teor. Fiz \textbf{71}, 738 (1976)

\bibitem{Lozovik_97} Y.E. Lozovik, O.L. Berman, J. Exp. and Th. Phys. \textbf{84}, 1027 (1997)

\bibitem{Rapaport_2012} Y. Shilo et al., Nat. Comm. \textbf{4}, 2335 (2013)

\bibitem{Alloing_2014} M. Alloing et al., Europhys. Lett. \textbf{107}, 10012 (2014); arXiv:1304.4101 (2013) 

\bibitem{Beian_2015} M. Beian et al., arXiv:1506.08020 (2015)

\bibitem{Ivanov_2010} A. L. Ivanov, E. A. Muljarov, L. Mouchliadis, and R. Zimmermann, Phys. Rev. Lett. \textbf{104}, 179701 (2010)

\bibitem{Schindler_2008} C. Schindler and R. Zimmermann, Phys. Rev. B \textbf{78}, 045313 (2008)

\bibitem{Timofeev_trap}	A. V. Gorbunov and V. B. Timofeev, JETP Lett. \textbf{84},  329 
(2006)

\bibitem{Butov_trap} A. A. High et al., Phys. Rev. Lett. \textbf{103}, 087403 (2009)

\bibitem{Butov_trap2} M. Remeika et al., Phys. Rev. Lett. \textbf{102}, 186803 (2009)

\bibitem{Holleitner_trap} G. Schinner et al., Phys. Rev. Lett. \textbf{110}, 127403 (2013)
	
\bibitem{Cohen_2016} K. Cohen et al., Nano Lett., \textbf{16}, 3726 (2016)

\bibitem{Stern_2014} M. Stern, V. Umansky and I. Bar-Joseph,  Science \textbf{343}, 55 (2014) 	
	
\bibitem{Eisenstein_2012} D. Nandi, A.D.K. Finck, J.P. Eisenstein, L.N. Pfeiffer and K.W. West, Nature \textbf{488}, 481 (2012)

\bibitem{Dietsche_2012} X. Huang, W. Dietsche, M. Hauser, and K. von Klitzing
Phys. Rev. Lett. \textbf{109}, 156802 (2012)

\bibitem{Ritchie_2008} A. F. Croxall et al., Phys. Rev. Lett. \textbf{101}, 246801 (2008)

\bibitem{Seamons_2009} J. A. Seamons, C. P. Morath, J. L. Reno, and M. P. Lilly
Phys. Rev. Lett. \textbf{102}, 026804 (2009)

\bibitem{Glauber_99} M. Naraschewski and R.J. Glauber, Phys. Rev. A \textbf{59}, 4595 (1999)

\bibitem{Berman_2004}O.L. Berman, Y.E. Lozovik, D.W. Snoke, and R.D. Coalson, Phys. Rev. B \textbf{70}, 235310 (2004)

\bibitem{Monique_2001} M. Combescot and C. Tanguy, EuroPhys. Lett. \textbf{55}, 390 (2001)

\bibitem{Bloch_2008} I. Bloch, J. Dalibard, and W. Zwerger, Rev. Mod. Phys. \textbf{80}, 885 (2008)


\bibitem{Weiler_2008} C. N. Weiler et al., Nature \textbf{455}, 948 (2008)

\bibitem{Navon_2015} N. Navon, A. L. Gaunt, R. P. Smith and Z. Hadzibabic, Science \textbf{347}, 167 (2015)

\bibitem{Zurek_96} W.H. Zurek, Phys. Rep. \textbf{276}, 177 (1996)




\end{thebibliography}

\begin{thebibliography}{99}

\bibitem{Beian_2015} M. Beian et al., arXiv:1506.08020 (2015)



\bibitem{High_2009} A. A. High et al., Phys. Rev. Lett. \textbf{103}, 087403 (2009)

\bibitem{Holleitner_2013} G. Schinner et al., Phys. Rev. Lett. \textbf{110}, 127403 (2013)

\bibitem{Ivanov_2010} A. L. Ivanov, E. A. Muljarov, L. Mouchliadis, and
R. Zimmermann, Phys. Rev. Lett. \textbf{104}, 179701 (2010)

\bibitem{Schindler_2008} C. Schindler and R. Zimmermann, Phys. Rev. B \textbf{78},
045313 (2008)



\bibitem{Blackwood} E. Blackwood et al., Phys. Rev. B \textbf{50}, 14246 (1994) 

\bibitem{Berman_2004}O.L. Berman, Y.E. Lozovik, D.W. Snoke, and R.D. Coalson, Phys. Rev. B \textbf{70}, 235310 (2004)

\end{thebibliography}
\end{document}